\documentclass[aps,pra,superscriptaddress,twocolumn,a4paper,reprint,eprint,longbibliography,amsmath,amsfonts,amssymb,footinbib]{revtex4-2}
\pdfoutput=1
\usepackage[caption = false]{subfig}
\usepackage{graphicx}
\usepackage{graphics}
\usepackage{xcolor}
\usepackage{bm}
\usepackage[percent]{overpic}
\usepackage{physics}
\usepackage[force]{feynmp-auto}
\usepackage[USenglish]{babel}
\usepackage{grffile}

\begin{document}
\title{Fate of the Hebel-Slichter peak in superconductors with strong antiferromagnetic fluctuations}
\author{D. C. Cavanagh}
\email{david.cavanagh@otago.ac.nz}
\affiliation{Department of Physics, University of Otago, P.O. Box
	56, Dunedin 9054, New Zealand}
\author{B. J. Powell}
\affiliation{School of Mathematics and Physics, The University of Queensland, Brisbane, Queensland 4072, Australia}
\begin{abstract}
We show that  magnetic fluctuations can destroy the Hebel-Slichter peak in conventional superconductors. The Hebel-Slichter peak has previously been expected to survive even in the presence of strong electronic interactions. However, we show that  antiferromagnetic fluctuations suppress the peak at $\bm{q}=0$ in the imaginary part of the magnetic susceptibility, $\chi_{+-}''\left(\bm{q},\omega\right)$, which causes the Hebel-Slichter peak. This  is of general interest as in many materials superconductivity is found near a magnetically ordered phase, and the absence of a Hebel-Slichter peak is taken as evidence of unconventional superconductivity in these systems. For example, no Hebel-Slichter peak is observed in the $\kappa$-(BEDT-TTF)$_2X$ organic superconductors but heat capacity measurements have been taken to indicate $s$-wave superconductivity. Similarly, experiments indicate nodeless superconductivity in many iron pnictide superconductors which exhibit no peak in the relaxation rate. If antiferromagnetic fluctuations destroy the putative Hebel-Slichter peak in organic superconductors then the peak should be restored by applying a pressure, which is known to suppress antiferromagnetic correlations in these materials.
\end{abstract}
\maketitle

\section{Introduction}

Unconventional superconductivity, and the identification of the underlying mechanism,  remains one of the most active areas of research in modern physics \cite{Mineev1999,Scalapino2012,Sigrist1991,Powell2006a,Norman2011}. The challenge of understanding unconventional superconductivity is compounded by the fact that macroscopic probes of the superconducting state are sensitive only to the emergent superconducting order parameter, or gap, and not to the microscopic mechanism responsible for it \cite{Annett1990,Annett1999,Powell2006,Wosnitza2007}. Any attempt to explain the microscopic origin of unconventional superconductivity must also explain how the resultant gap influences experiments.

Understanding the exact form of the superconducting gap is of considerable importance for developing an understanding of the microscopic basis for unconventional superconductivity, however, it is often far from straightforward in practice. The Josephson interference experiments responsible for unambiguously identifying the `d$_{x^2-y^2}$-wave' symmetry of the cuprates \cite{Wollman1993,Tsuei1994} have not been possible in many materials. The interpretation of other experimental results can be ambiguous, making a conclusive determination of the gap difficult. 

In many experiments, the bulk of the insight comes from the low temperature behavior of experimental probes, which reflect the density of states of the superconductor\cite{Tinkham2004,Mineev1999}. As such, the temperature dependence of these probes can be used to infer both the presence of nodes in the gap function and the form of these nodes (i.e. point or line nodes)\cite{Sigrist1991,Annett1990,Leggett1975}. Such probes cannot, however, identify the positioning of any nodes on the Fermi surface, and therefore cannot be used to differentiate between different gap functions with the same form of nodes. 

More detailed probes of the gap function are available for determining the presence and position of gap nodes on the Fermi surface. Such probes include the measurement of thermodynamic properties such as the heat capacity under a varying orientation of magnetic field \cite{Matsuda2006}, measurement of the structure of inelastic neutron scattering spectra \cite{Eremin2008}, as well as detailed surface probes such as scanning tunnelling spectroscopy and the related field of quasiparticle interference spectra probes \cite{Allan2013,Hirschfeld2015}. These experiments are often technically challenging, and many are only viable  for large samples and/or materials with extremely clean surfaces, which can considerably limit the applicability of such methods. 

One of the most notable probes of the superconducting gap is the spin-lattice relaxation rate $1/T_1$ measured in nuclear magnetic resonance. For conventional superconductors,  $1/T_1$ displays a peak below the superconducting transition temperature known as the Hebel-Slichter peak \cite{Hebel1957,Hebel1959}. The existence of this peak was one of the earliest confirmations of the Bardeen-Cooper-Schrieffer theory of conventional superconductivity \cite{Bardeen1957,Bardeen1957a}, indicating the presence of a coherent state. Experimentally, the presence of such a peak has long been taken as a key signature of superconductivity with an isotropic, nodeless gap. Here, by investigating general features of the behavior in simple model systems, we seek to understand the influence of magnetic fluctuations on $1/T_1$ in general, and the form of the Hebel-Slichter peak in particular.

In the most interesting superconductors, both strong antiferromagnetic fluctuations and the absence of a Hebel-Slichter peak in $1/T_1$ are ubiquitous \cite{Scalapino2012,Annett1999,Powell2011}. Additionally, the form of the gap function in many such materials remains contentious. One  significant example is the  organic superconductor, $\kappa$-(BEDT-TTF)$_2$Cu[N(CN)$_2$]Br ($\kappa$-Br). This material has the highest critical temperature (at ambient pressure) of the BEDT-TTF based superconductors \cite{Wosnitza2007}, $\kappa$-Br has been subjected to a wide variety of experimental probes over the last three decades. Despite this, the symmetry of the superconducting order parameter in this material remains a matter of considerable disagreement \cite{Schmalian1998, Kuroki2002, Powell2006, Powell2006a, Powell2004, Powell2007, Guterding2016a, Guterding2016, Zantout2018, Wosnitza2003, Elsinger2000,  Kuehlmorgen2017, Milbradt2013,Wosnitza2012}. Strong antiferromagnetic fluctuations are observed in $\kappa$-Br \cite{Yusuf2007,Powell2009}; indeed the closely related material $\kappa$-(BEDT-TTF)$_2$Cu[N(CN)$_2$]Cl  ($\kappa$-Cl) is an antiferromagnetic insulator at ambient pressure that can be driven superconducting by moderate hydrostatic pressures \cite{Powell2006a}. The superconducting states \cite{Powell2009} and magnetic fluctuations \cite{Powell2006a} in metallic/superconducting $\kappa$-Cl and $\kappa$-Br are extremely similar.

While Knight shift measurements on $\kappa$-Br consistently indicate singlet pairing \cite{Mayaffre1995,Kanoda1996}, due to the vanishing of the Knight shift at zero temperature, interpretations of the results of other experiments have been inconsistent. There has been evidence from the temperature dependence of low temperature specific heat measurements to indicate nodeless (`s-wave') superconductivity \cite{Elsinger2000,Mueller2002,Wosnitza2003} while other experiments indicate the presence of nodes of the gap function \cite{Nakazawa1997,Taylor2007,Malone2010}. Similarly,  penetration depth measurements were contentious \cite{Lang1992,Yoneyama2004,Carrington1999} until recently, with more precise measurements showing a power law temperature dependence suggestive of a nodal superconducting state \cite{Milbradt2013}. The density of states from the surface tunneling spectroscopy has been found to show some indication of multiple coherence peaks, which has been interpreted in terms of a complicated mixed order parameter \cite{Guterding2016}, although similar signals have been observed in other multi-band materials with a superconducting gap magnitude that varies between bands \cite{Giubileo2002}. This ongoing disagreement has led both theorists \cite{Schmalian1998,Powell2005,Powell2006,Powell2007,Guterding2016a,Cavanagh2018} and experimentalists \cite{Dion2009,Guterding2016} to discuss the possibility of a variety of superconducting gaps, including those with symmetry required or accidental nodes in organic superconductors. 

Despite the lack of an observed Hebel-Slichter peak in $1/T_1$ \cite{Mayaffre1995,Kanoda1996}, there are some who argue that the superconducting gap may in fact be nodeless \cite{Wosnitza2007,Wosnitza2012}, as supported by recent thermal conductivity measurements  \cite{Kuehlmorgen2017}. In such a scenario, the absence of a Hebel-Slichter peak needs a detailed explanation. Thus, it is of significant interest to understand how magnetic fluctuations influence the $1/T_1$ relaxation rate and whether the suppression of a peak by spin-fluctuations is sufficient to explain the relaxation rate in such materials. In particular, we will focus on antiferromagnetic fluctuations described by the random phase approximation (RPA), and how their relative strength influences the relaxation rate, with a focus on the effects on the Hebel-Slichter peak. 

Early attempts to understand the unconventional superconductivity in the cuprates, found that coherence effects can, in principle, be disguised by a combination of strong-coupling and electronic interactions \cite{Statt1990,Akis1991a}. It was also found, however, that these effects alone were insufficient to match experimental data with an isotropic gap \cite{Bulut1992,Bulut1992a}. However, a detailed analysis of the influence of magnetic fluctuations on the Hebel-Slichter peak is currently lacking. For example, the absence of the Hebel-Slichter peak, and the potential role played by magnetic fluctuations has not previously been examined in the organic superconductors.
 
These materials are of particular interest because the bandwidth, and therefore the relative strength of electronic interactions, in these materials is tunable by the application of external pressure \cite{Wosnitza2007, Dumm2009}. Therefore, it may be possible to alter the interaction strength and determine the gap structure by measuring $1/T_1T$ and comparing both the temperature and interaction dependence of the relaxation rate to predictions. We show that in these materials the suppression of the Hebel-Slichter peak can in principle be explained  entirely due to the influence of spin fluctuations, rather than gapless superconductivity, as has been discussed previously \cite{Koyama1989}.

We additionally consider a model of the iron pnictide superconductors, a large family of complex materials with strong spin-orbit coupling and multiple bands at the Fermi level. While various superconducting gaps have been proposed \cite{Raghu2008, Cvetkovic2009, Vafek2017, Ong2016}, the majority of experiments support a $s_{\pm}$-wave superconducting state \cite{Chubukov2008, Mazin2008, Paglione2010, Chubukov2012}, which is relatively isotropic on the Fermi surface sheets, but changes sign between bands. These materials are known to have strong spin-fluctuations and exhibit no Hebel-Slichter peak in $1/T_1T$ despite the presence of a superconducting gap that most likely transforms as the trivial ($A_{1g}$) representation.

In a previous work \cite{Cavanagh2018}, we demonstrated the potential use of the nuclear magnetic relaxation rate, $1/T_1T$, to experimentally differentiate between those gaps with accidental nodes (i.e. nodes not required by symmetry) and those gaps with nodal positions constrained by symmetry, due to a peak arising in $1/T_1T$ for the former case immediately below $T_c$, similar to the well known Hebel-Slichter peak found in nodeless superconductors. In addition to considering the Hebel-Slichter peak in isotropic superconductors, we will also address the suppression of this Hebel-Slichter-like peak by antiferromagnetic fluctuations.

\section{Theory}

The spin lattice relaxation rate, $1/T_1$, measured in nuclear magnetic resonance, is related to the transverse spin susceptibility, $\chi_{+-}\left(\bm{q},\omega\right)=\chi_{+-}'\left(\bm{q},\omega\right)+i\chi_{+-}''\left(\bm{q},\omega\right)$, via
\begin{eqnarray}
\frac{1}{T_1T}&=& \lim\limits_{\omega\rightarrow 0} \frac{2k_B}{\gamma_e^2\hbar^4}\sum\limits_{\bm{q}}\left|A_H(\bm{q})\right|^2\frac{\chi_{+-}''\left(\bm{q},\omega\right)}{\omega},\label{T1T_def}
\end{eqnarray}
where $\gamma_e$ is the electronic gyromagnetic ratio, $A_H(\bm{q})$ is the hyperfine coupling, which we will approximate by a point contact interaction, constant with respect to $\bm{q}$. In a conventional nodeless superconductor, the relaxation rate increases below $T_c$ to a peak before decreasing rapidly as temperature is lowered. Formally, the peak arises due to a divergence in the relaxation rate that is cut off by a combination of effects due to impurities, slight anisotropy of the gap, electronic interactions or in the extreme limit, by the influence of the crystal lattice, which sets a characteristic length scale \cite{Ketterson1999,Tinkham2004}. The fact that such a divergence is absent in the majority of unconventional superconductors is typically taken as evidence of nodes in the superconducting gap \cite{Bulut1992,Scalapino1995,Mineev1999}, though in some cases it has been argued that strong electronic correlations may be responsible for the suppression of the peak \cite{Wosnitza2007}. 
The magnetic susceptibility in the superconducting state, in the absence of vertex corrections is given by,
\begin{widetext}
\begin{eqnarray}
\chi_{+-}\left(\bm{q},\omega\right)=\chi_0\left(\bm{q},\omega\right) &=& \frac{1}{N}\sum\limits_{\bm{k}} \left\lbrace \frac{1}{2}\left[1+\frac{\xi_{\bm{k}+\bm{q}}\xi_{\bm{k}}+\Delta_{\bm{k}+\bm{q}}\Delta_{\bm{k}}}{E_{\bm{k}+\bm{q}}E_{\bm{k}}}\right]\frac{f\left(E_{\bm{k}+\bm{q}}\right)-f\left(E_{\bm{k}}\right)}{\omega - \left(E_{\bm{k}+\bm{q}}-E_{\bm{k}}\right)+i\eta}+ \frac{1}{4}\left[1-\frac{\xi_{\bm{k}+\bm{q}}\xi_{\bm{k}}+\Delta_{\bm{k}+\bm{q}}\Delta_{\bm{k}}}{E_{\bm{k}+\bm{q}}E_{\bm{k}}}\right]\right.\nonumber\\
&& \left.\times\frac{\bar{f}\left(E_{\bm{k}+\bm{q}}\right)-f\left(E_{\bm{k}}\right)}{\omega + \left(E_{\bm{k}+\bm{q}}+E_{\bm{k}}\right)+i\eta}+\frac{1}{4}\left[1-\frac{\xi_{\bm{k}+\bm{q}}\xi_{\bm{k}}+\Delta_{\bm{k}+\bm{q}}\Delta_{\bm{k}}}{E_{\bm{k}+\bm{q}}E_{\bm{k}}}\right]\frac{f\left(E_{\bm{k}+\bm{q}}\right)-\bar{f}\left(E_{\bm{k}}\right)}{\omega - \left(E_{\bm{k}+\bm{q}}+E_{\bm{k}}\right)+i\eta}\right\rbrace ,
\end{eqnarray}
\end{widetext}
where $E_{\bm{k}}=\sqrt{\xi^2_{\bm{k}}+\Delta^2_{\bm{k}}}$ is the superconducting quasiparticle energy, defined in terms of the electron dispersion $\xi_{\bm{k}}=\varepsilon_{\bm{k}}-\mu$ and the superconducting gap $\Delta_{\bm{k}}$, $f\left(E\right)$ is the Fermi-Dirac distribution function [$\bar{f}\left(E\right)=1-f\left(E\right)$], and in the absence of interactions the limit of the lifetime $\eta\rightarrow 0^+$ is implied. 

\subsection{Anisotropic gaps with accidental nodes}
Previously \cite{Cavanagh2018}, we demonstrated the possibility of a Hebel-Slichter-like peak emerging in the relaxation rate in systems where the superconducting gap is nodal, but the location of the nodes is not dictated by symmetry. In such systems, the (in general) nonzero average of the gap over the Fermi surface gives rise to a peak in the relaxation rate analogous to the Hebel-Slichter peak, even if the integral of the superconducting gap  over the Brillouin zone vanishes.

Lifetime effects (via the self-energy) on $1/T_1T$ have already been investigated to a degree in Ref. \cite{Cavanagh2018}, where a finite quasiparticle lifetime was introduced into the numerical calculations. This served the purpose of investigating the contribution of impurity effects on the Hebel-Slichter-like peaks. Including electronic interactions in the quasiparticle lifetime is not expected to alter the picture dramatically, introducing a temperature dependence to the lifetime but not significantly influencing the stability of the peak structure \cite{Cavanagh2018}. In this work, we investigate the effects of antiferromagnetic fluctuations and show that they have a much more dramatic effect.

\subsection{The Random Phase Approximation}

In the absence of vertex corrections, the transverse spin susceptibility can be expressed in terms of a convolution of single-particle propagators,
\begin{widetext}
\begin{eqnarray}
\chi_{+-}\left(\bm{q},\omega\right)=\lim\limits_{i\omega_n\rightarrow \omega +i\eta}\sum\limits_{\bm{k},i\Omega_m\bar{\sigma}\ne\sigma}G^{(0)}_{\bm{k}+\bm{q},\sigma} \left(i\omega_n\right)G^{(0)}_{\bm{k},\bar{\sigma}} \left(i\omega_n +i\Omega_m\right) ,\label{chipm}
\end{eqnarray}
in which case the relaxation rate can be expressed with the influence of the two Green's functions separated \cite{Cavanagh2018}, due to a property of the convolution, $\sum_{\bm{q},\bm{k}} f\left(\bm{k}+\bm{q}\right)f\left(\bm{k}\right)=\left[\sum_{\bm{k}} f\left(\bm{k}\right)\right]^2$, and
\begin{eqnarray}
\frac{1}{T_1T}  \propto\int d E \left(-\frac{df}{dE}\right)\left\lbrace\left[\sum\limits_{\bm{k}}A_{\bm{k},\bar{\sigma}} \left(E\right)\right]^2+\left[\sum\limits_{\bm{k}}\frac{\xi_{\bm{k}}}{E}A_{\bm{k},\bar{\sigma}} \left(E\right)\right]^2+\left[\sum\limits_{\bm{k}}\frac{\Delta_{\bm{k}}}{E}A_{\bm{k},\bar{\sigma}} \left(E\right)\right]^2\right\rbrace,\label{eq:noVC}
\end{eqnarray}
\end{widetext}
where $A_{\bm{k},\bar{\sigma}} \left(E\right)$ is the quasiparticle spectral density function. The last two terms in Eq. \ref{eq:noVC} arise from the coherence factors of the Green's functions and represent a Fermi surface average of the dispersion and superconducting gap, the later of which is the origin of the Hebel-Slichter peak and the related peak in superconductors with accidental nodes \cite{Cavanagh2018}. In a superconductor with an isotropic gap, the Hebel-Slichter peak may be more intuitively understood as originating from the coherence peak in the density of states \cite{Tinkham2004}, but this explanation is not sufficient to explain the presence of a similar peak for a gap with accidental nodes which is not present for anisotropic gaps with symmetry protected nodes. 

The simplified expression \ref{eq:noVC} is valid only when vertex corrections, in particular those due to spin-fluctuations, are absent. In this limit, the susceptibility, $\chi_{+-}\left(\bm{q},\omega\right)$, depends on the momentum $\bm{q}$ solely through the spectral density functions. In the presence of strong spin fluctuations, the susceptibility can no longer be written in this way, and must take account of the influence of the spin fluctuations. In order to investigate these effects, we turn to the random phase approximation, as the simplest treatment. As it is not our intention to discuss the influence of the spin fluctuations on the superconductivity itself, we refrain from the use of more computationally expensive self-consistent methods (such as the fluctuation-exchange, or FLEX, approximation). While treating spin fluctuations via the RPA alone is not sufficient to ensure thermodynamic consistency \cite{Vilk1997}, this approach is standard, and sufficient to describe the leading order effects of spin fluctuations on the NMR relaxation rate \cite{Yusuf2007,Yusuf2009,Bulut1992,Bulut1992a,Kobayashi1999}.

Within the RPA, the susceptibility is given by a sum over` ladder diagrams \cite{Doniach1998},
\begin{widetext}
\begin{align}
\chi_{\mathrm{RPA}}\left(\bm{q},\omega\right)&=
\parbox{20mm}{
\begin{fmffile}{rpat}
        \begin{fmfchar*}(50,25)
            \fmfleft{i}
            \fmfright{o}
                \fmftop{v1}
            \fmfbottom{v2}
            \fmf{phantom,left=0.4,label=$\uparrow$}{i,o}
            \fmf{phantom,left=0.4,label=$\downarrow$}{o,i}
               \fmffreeze
            \fmf{phantom}{i,G3}
            \fmfpoly{shaded,tension=0.01,label=$\hspace{1.2cm}\Gamma_\textrm{RPA}$}{G3,G2,G1}
            \fmf{fermion,left=0.3,tension=0.01}{G1,o}
            \fmf{fermion,left=0.3,tension=0.01}{o,G2}
        \end{fmfchar*}
    \end{fmffile}
}\\
 \nonumber\\
    \nonumber\\
    &=
\parbox{20mm}{
	\begin{fmffile}{rpa0}
        \begin{fmfchar*}(50,25)
            \fmfleft{i}
            \fmfright{o}
                \fmftop{v1}
            \fmfbottom{v2}
            \fmf{fermion,left=0.4,label=$\uparrow$}{i,o}
            \fmf{fermion,left=0.4,label=$\downarrow$}{o,i}
        \end{fmfchar*}
    \end{fmffile}
 } +\text{  }\parbox{20mm}{
\begin{fmffile}{rpa1}
        \begin{fmfchar*}(50,25)
            \fmfleft{i}
            \fmfright{o}
                \fmftop{v1}
            \fmfbottom{v2}
                \fmf{fermion,left=0.2}{i,v1,o}
            \fmf{fermion,left=0.2}{o,v2,i}
            \fmf{dashes,label=$U$,l.d=.03w}{v1,v2}
               \fmflabel{$\uparrow$}{v1}
               \fmflabel{$\downarrow$}{v2}
        \end{fmfchar*}
    \end{fmffile}
} +\text{  }\parbox{20mm}{
\begin{fmffile}{rpa2}
        \begin{fmfchar*}(50,25)
            \fmfleft{i}
            \fmfright{o}
                \fmftop{x1,w1,v1,u1}
            \fmfbottom{x2,w2,v2,u2}
                \fmf{fermion,left=0.1,label=$\uparrow$}{w1,v1}
            \fmf{fermion,left=0.1,label=$\downarrow$}{v2,w2}
                \fmf{fermion,left=0.2}{i,w1}
            \fmf{fermion,left=0.2}{o,v2}
                \fmf{fermion,left=0.2}{v1,o}
            \fmf{fermion,left=0.2}{w2,i}
            \fmf{dashes,label=$U$,l.d=.03w}{v1,v2}
            \fmf{dashes,label=$U$,l.d=.03w}{w1,w2}
        \end{fmfchar*}
    \end{fmffile}}+\dots\nonumber\\
    \nonumber\\
&= \frac{\chi_{0}\left(\bm{q},\omega\right)}{1-U\chi_{0}\left(\bm{q},\omega\right)},
\label{eq:rpa_diag}
\end{align}
\end{widetext}
where $\chi_{0}\left(\bm{q},\omega\right)$ is the bare transverse magnetic (superconducting) susceptibility, and $U$ is a Hubbard interaction parameter (longer range interactions  introduce a momentum dependence in this interaction parameter). 
The ladder diagrams here are in marked contrast with the bubble diagrams that emerge in the RPA for the contributions of the long-range Coulomb interaction to the dielectric function \cite{Mahan2000}.

The imaginary part of the susceptibility, which enters into $1/T_1T$, is then given by
\begin{eqnarray}
\chi''_{RPA}\left(\bm{q},\omega\right) &=& \frac{\chi''_{0}\left(\bm{q},\omega\right)}{\left[1-U\chi'_{0}\left(\bm{q},\omega\right)\right]^2+\left[U\chi''_{0}\left(\bm{q},\omega\right)\right]^2}.
\end{eqnarray}
Within the framework of the RPA, the transition to a magnetically ordered state is described by a divergence in the static ($\omega =0$) susceptibility. The real and imaginary parts of the susceptibility are related by a Kramers-Kronig transformation, as a result of a fluctuation-dissipation theorem \cite{Giuliani2005,Coleman2015,Arfken2005}. One of the consequences of this relationship is that, at low frequencies, the imaginary part of the susceptibility varies linearly with frequency, vanishing in the static limit, while the real part tends to a constant value. The divergence of the susceptibility in the RPA then must occur when $U\chi'\left(\bm{q},0\right)=1$. This corresponds to a magnetic instability in the material and the RPA predicts long-range antiferromagnetic order for $U>U_c$. Thus, $U_c=1/\text{max}\left[\chi'\left(\bm{q},0\right)\right]$ sets an upper limit for the interaction strength in numerical calculations. Understanding the effects of the spin fluctuations on superconductivity and possible phase transitions near $U_c$, is not the focus of this work, and would require a more sophisticated self-consistent approach. As such, we have chosen not to address this question at this time, but rather focus on the universal behavior of the Hebel-Slichter peak in the presence of spin-fluctuations, which can be demonstrated straightforwardly in the RPA.

\subsubsection{Effects on the $1/T_1$ relaxation rate}
To fully understand the effects of spin fluctuations on $1/T_1T$, it is necessary to resort to numerical calculations (see Section \ref{num}), but some insight may still be gained analytically. In the low frequency limit, the susceptibility may be approximated by
\begin{align}
\chi_{0}' \left(\bm{q},\omega\right)&\approx \chi_{0}' \left(\bm{q},0\right)\equiv B_{\bm{q}}\left(T\right) , \nonumber\\
\frac{\chi_{0}''\left(\bm{q},\omega\right)}{\omega}&\approx \left.\frac{\chi_{0}''\left(\bm{q},\omega\right)}{\omega}\right|_{\omega\rightarrow 0} \equiv C_{\bm{q}}\left(T\right), \nonumber\\
\end{align}
in which case the relaxation rate is given by
\begin{eqnarray}
\frac{1}{T_1T}&\propto & \lim\limits_{\omega\rightarrow 0} \sum\limits_{\bm{q}}\frac{1}{\omega}\frac{\chi_{0}''\left(\bm{q},\omega\right)}{\left[1-U\chi_{0}'\left(\bm{q},\omega\right)\right]^2+\left[U\chi_{0}''\left(\bm{q},\omega\right)\right]^2}\nonumber\\
&=& \lim\limits_{\omega\rightarrow 0} \sum\limits_{\bm{q}}\frac{C_{\bm{q}}\left(T\right)}{\left[1-U B_{\bm{q}}\left(T\right)\right]^2+\left[UC_{\bm{q}}\left(T\right)\omega\right]^2}\nonumber\\
&= &\sum\limits_{\bm{q}}\frac{C_{\bm{q}}\left(T\right)}{\left[1-U B_{\bm{q}}\left(T\right)\right]^2}.
\end{eqnarray}
In the absence of antiferromagnetic fluctuations the relevant features are given by the form of $C_{\bm{q}}$, which are influenced, when $U\neq 0$, by features of the static real part of the susceptibility. In particular, since $UB_{\bm{q}}\leq 1$, whenever $B_{\bm{q}}\approx 1$, the contribution to the relaxation rate is enhanced, and when $B_{\bm{q}}$ is small, features of $C_{\bm{q}}$ are suppressed. 

The static part of the susceptibility in a superconductor is given by
\begin{widetext}
\begin{align}
B_{\bm{q}}\left(T\right)&=\chi'_{0}\left(\bm{q},\omega=0\right)\nonumber\\
&=\sum\limits_{\bm{k}}\left\lbrace \frac{1}{2}\left[1+\frac{\xi_{\bm{k}}\xi_{\bm{k}+\bm{q}}+\Delta_{\bm{k}}\Delta_{\bm{k}+\bm{q}}}{E_{\bm{k}}E_{\bm{k}+\bm{q}}}\right]\frac{f\left( E_{\bm{k}+\bm{q}}\right)-f\left(E_{\bm{k}}\right)}{ E_{\bm{k}} - E_{\bm{k}+\bm{q}}}+\frac{1}{4}\left[1-\frac{\xi_{\bm{k}}\xi_{\bm{k}+\bm{q}}+\Delta_{\bm{k}}\Delta_{\bm{k}+\bm{q}}}{E_{\bm{k}}E_{\bm{k}+\bm{q}}}\right]\frac{\bar{f}\left( E_{\bm{k}+\bm{q}}\right)-f\left(E_{\bm{k}}\right)}{E_{\bm{k}} + E_{\bm{k}+\bm{q}}}\right. \nonumber\\
&\qquad \qquad \left. +\frac{1}{4}\left[1-\frac{\xi_{\bm{k}}\xi_{\bm{k}+\bm{q}}+\Delta_{\bm{k}}\Delta_{\bm{k}+\bm{q}}}{E_{\bm{k}}E_{\bm{k}+\bm{q}}}\right]\frac{\bar{f}\left(E_{\bm{k}}\right)-f\left( E_{\bm{k}+\bm{q}}\right)}{E_{\bm{k}} + E_{\bm{k}+\bm{q}}}\right\rbrace , \label{staticChi}
\end{align}
\end{widetext}
and the structure of $B_{\bm{q}}$ can be seen to depend in a complicated manner on the band structure and gap symmetry, particularly with regards to approximate nesting of the Fermi surface (which may enhance the first term). 

The real part of the susceptibility can, in principle, enhance the features dominating the relaxation rate (i.e. if $C_{\bm{q}}\left(T\right)$ and $B_{\bm{q}}\left(T\right)$ have similar momentum-dependence, large contributions to the relaxation rate will be enhanced while smaller contributions will be unaffected). There is, however, no \emph{a priori} reason to expect such enhancement, as the momentum-dependences of $C_{\bm{q}}\left(T\right)$ and $B_{\bm{q}}\left(T\right)$ may differ drastically. 

To gain further insight into the influence of antiferromagnetic fluctuations on the relaxation rate, it becomes necessary to turn to specific models and numerical calculations, as is the focus of the remainder of this work.

\section{Numerical Results}\label{num}
In order to explore the behavior of $1/T_1T$ in the presence of antiferromagnetic spin fluctuations, we numerically calculate the relaxation rate for various interaction strengths. 

\subsection{Effective models}
To highlight the generality of our results, we consider three concrete examples. The first two models specifically include strong anisotropy and in both cases consider fully gapped superconducting states and those with accidental nodes. As a final example, we consider a model iron pnictide superconductor, a system for which spin-fluctuations are known to be strong and the gap is believed to belong to the trivial representation despite the lack of a Hebel-Slichter peak in $1/T_1T$ \cite{Oka2012}.

The first example is a toy model with anisotropic hopping parameters along the two axes,
\begin{equation}
\xi_{\bm{k}}=t_x\cos\left(k_x\right)+t_y\cos\left(k_y\right).
\end{equation}
Such a model is useful in demonstrating effects arising in a d-wave superconducting state with accidental nodes. For example, for $t_x\ne t_y$ the nodes in a superconducting gap with
\begin{equation}
\Delta_{\bm{k}}=\frac{\Delta_0}{2}\left[\cos\left(k_x\right)-\cos\left(k_y\right)\right],
\end{equation}
are not symmetry required (for example, adding a small s-wave component does not change the symmetry or cause a phase transition) and (even without an $s$-wave component) the average of the order parameter over the Fermi surface is non-zero \cite{Cavanagh2018}. 

\begin{figure}
	\centering
	\includegraphics[width=0.45\textwidth]{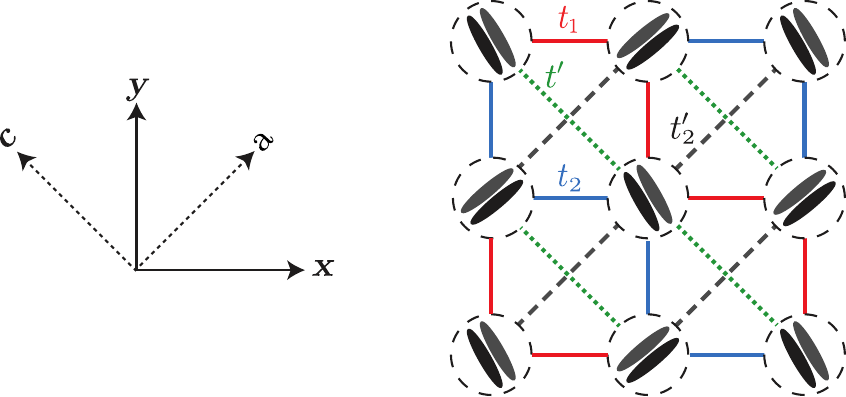}
	\caption{Representation of the tight binding model for $\kappa$-Br, as defined in Ref. \onlinecite{Koretsune2014}, including anisotropic next-nearest-neighbor hopping parameters, $\left|t'_2\right|<\left|t'\right|$. Here $a$  and $c$ here are crystallographic axes for the orthorhombic unit cell of $\kappa$-(BEDT-TTF)$_2$Cu[N(CN)$_2$]Br, with the model axes $x$ and $y$ rotated by 45 degrees. The unit cell is indicated by the gray area.}
	\label{OrganicModel}
\end{figure}

The second model we consider is a two-band effective tight-binding model for $\kappa$-Br, with hopping magnitudes parametrized by density functional theory \cite{Koretsune2014}. This model offers the opportunity to understand the resilience of the Hebel-Slichter peak in a more realistic band structure. Additionally, the $\kappa$-Br model allows us to make comparison between several proposed gap functions for the material. In this model, the BEDT-TTF dimers are treated as sites, and the tight-binding parameters are displayed schematically in Fig. \ref{OrganicModel}, with dispersion given by
\begin{widetext}
	\begin{equation}
	\xi_{\bm{k},\pm} = t'\cos\left(k_c\right) +t_2'\cos\left(k_a\right) \pm t\sqrt{\left[\cos\left(\frac{k_a+k_c}{2}\right)+\cos\left(\frac{k_a-k_c}{2}\right)\right]^2+\left(\frac{\delta_t}{t}\right)^2\left[\sin\left(\frac{k_a+k_c}{2}\right)+\sin\left(\frac{k_a-k_c}{2}\right)\right]^2}, \label{HottaDimerEq}
	\end{equation}
\end{widetext}
where $t'$ and $t'_2$ are (anisotropic) hopping parameters between next-nearest-neighbor dimers, $t=\left(t_1+t_2\right)/2$ is the average hopping along the $x$ and $y$ directions, and $\delta_t=\left(t_1-t_2\right)/2$ the difference between the alternating hopping strengths (which are dependent on the dimer orientation).

Due to the anisotropy of this model, the `d$_{xy}$-wave' state has accidental nodes, giving rise to a Hebel-Slichter-like peak in $1/T_1T$, and is given by
\begin{equation}
\Delta_{\bm{k}}^{\left(\text{x}\text{y}\right)} = \Delta_0 \sin\left(k_x\right)\sin\left(k_y\right).\label{accGap_K}
\end{equation}

Finally, for the iron pnictide superconductors we use a simple two-orbital model, as  proposed in Ref. \cite{Raghu2008} with an additional spin-orbit coupling $\lambda$ allowed by symmetry and in keeping with more general models \cite{Cvetkovic2013, Vafek2017}, with dispersion
\begin{widetext}
\begin{equation}
	\xi_{\bm{k},\pm} = \left(t_1+t_2\right)\left[\cos\left(k_x\right) +\cos\left(k_y\right)\right] + 4t_3\cos\left(k_x\right)\cos\left(k_y\right) \pm \sqrt{\left(t_1-t_2\right)^2\left[\cos\left(k_x\right) -\cos\left(k_y\right)\right]^2 + \left[2t_3\sin\left(k_x\right)\sin\left(k_y\right)\right]^2 +\lambda^2}, \label{Raghu_disp}
\end{equation}
\end{widetext}
and we take $(t_1,t_2,t_3,\lambda)=(-1,1.3,-0.85,0.1)|t_1|$ with $\mu=1.45|t_1|$. For brevity, we consider only a simple $s_{\pm}$-wave gap, which is the most common used for analysis of the iron pnictide materials,
\begin{equation}
\Delta_{\bm{k}} = \Delta_0 \cos\left(k_x\right)\cos\left(k_y\right).\label{sPM_gap}
\end{equation}
We note that, while here we do not consider the possibility of accidental nodes in the iron pnictide model, there have been reports of nodal superconductivity in some regimes for these materials \cite{Ong2016, Okazaki2012, Okazaki2012PRL, Watanabe2014, Abdiel2013}.

\subsection{Results}

\begin{figure}
\centering
		\begin{overpic}[width=0.45\textwidth]{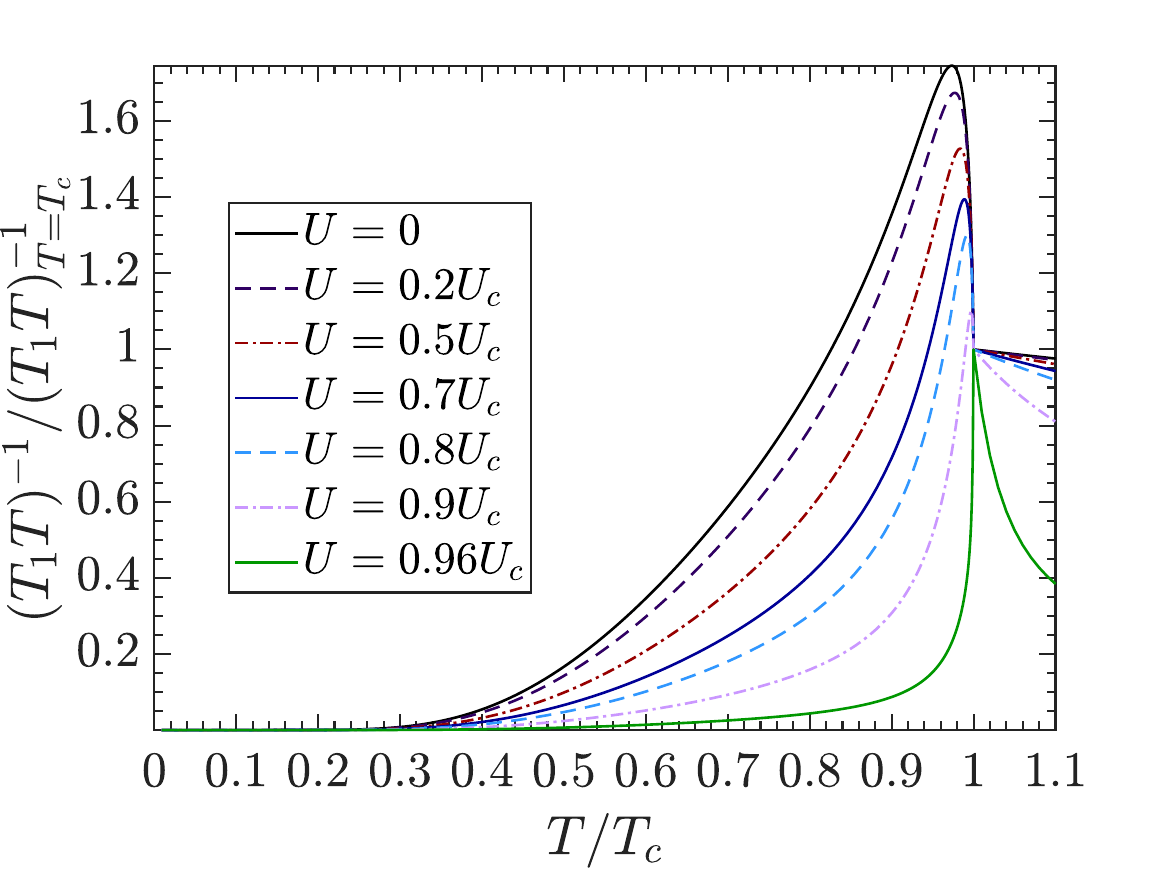}
		\put (0,70) {\bf a)}
		\end{overpic} \\
		\begin{overpic}[width=0.45\textwidth]{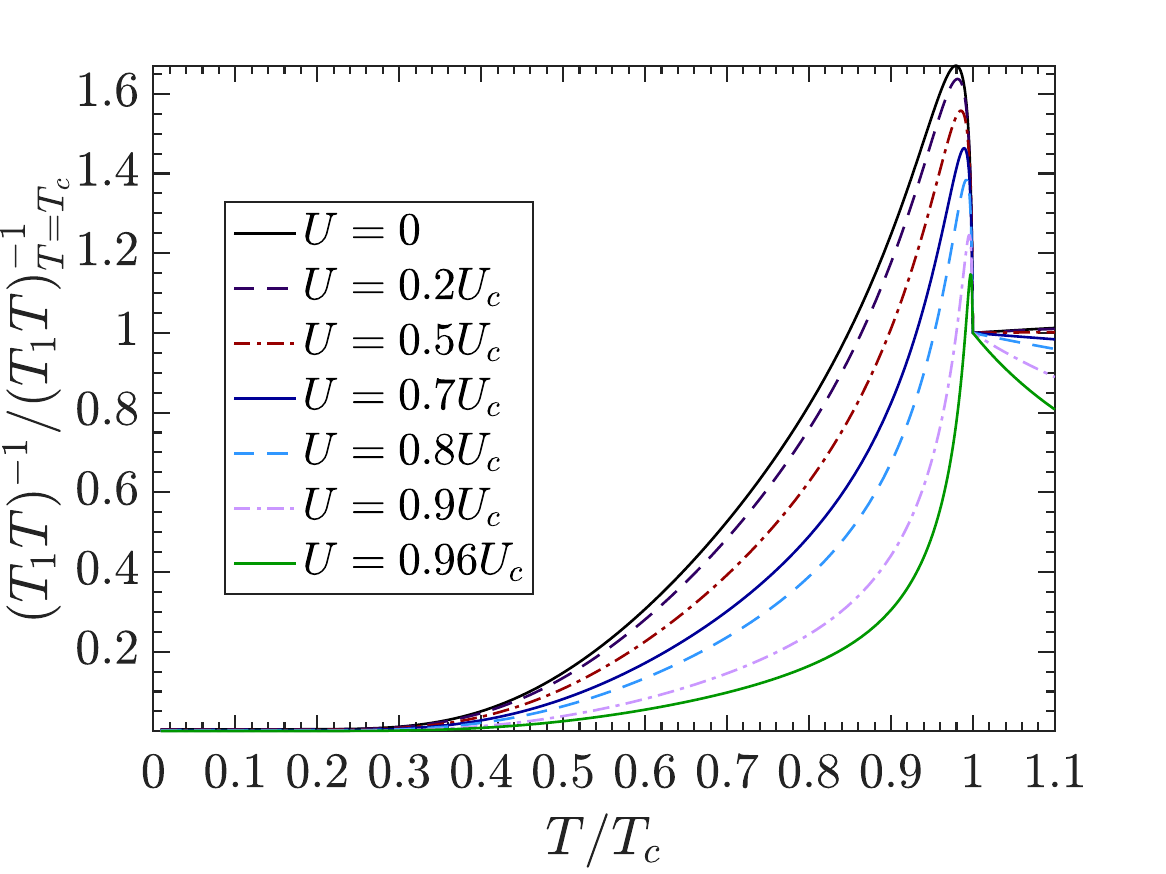}
		\put (0,70) {\bf b)}
		\end{overpic}
	\caption{The temperature dependence of $1/T_1T$, and suppression of the Hebel-Slichter peak, with increasing interaction strength in (a) the orthorhombic model with $t_x=0.4t_y$, and (b) a model for $\kappa$-Br, with an isotropic nodeless $s$-wave gap. In the limit of strong antiferromagnetic fluctuations, the peak narrows and eventually vanishes entirely. While the peak only disappears for $U\geq 0.95U_c$, the narrowing and suppression of the peak at lower interaction strengths may be sufficient to disguise the Hebel-Slichter peak in experiments. At low temperatures, $1/T_1T$ has an exponential temperature dependence, even in the presence of strong antiferromagnetic fluctuations. In these calculations, $\eta = 5\times 10^{-3} t$, $\omega =  10^{-3} t$, and $\Delta_0/2=2.5k_BT_c$, while $U_c\sim 11.5t$ for the orthorhombic model and $U_c\sim 9.6t$ for the model of $\kappa$-Br. }
	\label{U_sRPA}
\end{figure}

The suppression of the Hebel-Slichter peak is shown for a purely isotropic s-wave gap function in Fig. \ref{U_sRPA}, for both the orthorhombic model with $t_y=0.4t_x$ at quarter filling and for the effective model of $\kappa$-Br, with $\left(t',t_2',\delta_t\right)=\left(-0.54,0.14,0.03\right)t$. The parameters for the orthorhombic model are chosen to maximize the anisotropy of the Fermi surface while ensuring there are no Van Hove singularities close to the Fermi energy, while the model of $\kappa$-Br has been parametrized from first-principles calculations by Ref. \cite{Koretsune2014}. In both cases, we found that variation of the model parameters had little influence on the resulting relaxation rate. As the interaction strength increases, the prominent Hebel-Slichter peak is gradually reduced in magnitude and narrows, until the peak finally vanishes for both models near the phase transition to long-range magnetic order ($U\geq 0.95U_c$). While the peak is absent entirely only very close to the antiferromagnetic instability, the narrowing of the peak may be sufficient in some experiments to disguise its presence, depending on the temperature resolution of the experiment. 

At low temperatures, the increasing interaction strength reduces the overall magnitude of the relaxation rate but does not alter the temperature dependence, which displays the exponential suppression of quasiparticle states at low energies. The Hebel-Slichter peak is noticeably more resilient to the strength of the spin fluctuations for the effective model of $\kappa$-Br than the toy orthorhombic model, and the absence of any such peak in experiments \cite{Kanoda1996} is therefore inconsistent with an isotropic gap even for strong interactions, in contrast with interpretations of some other experiments \cite{Elsinger2000,Kuehlmorgen2017}.

\begin{figure*}
	\centering 
		\begin{overpic}[width=0.45\textwidth]{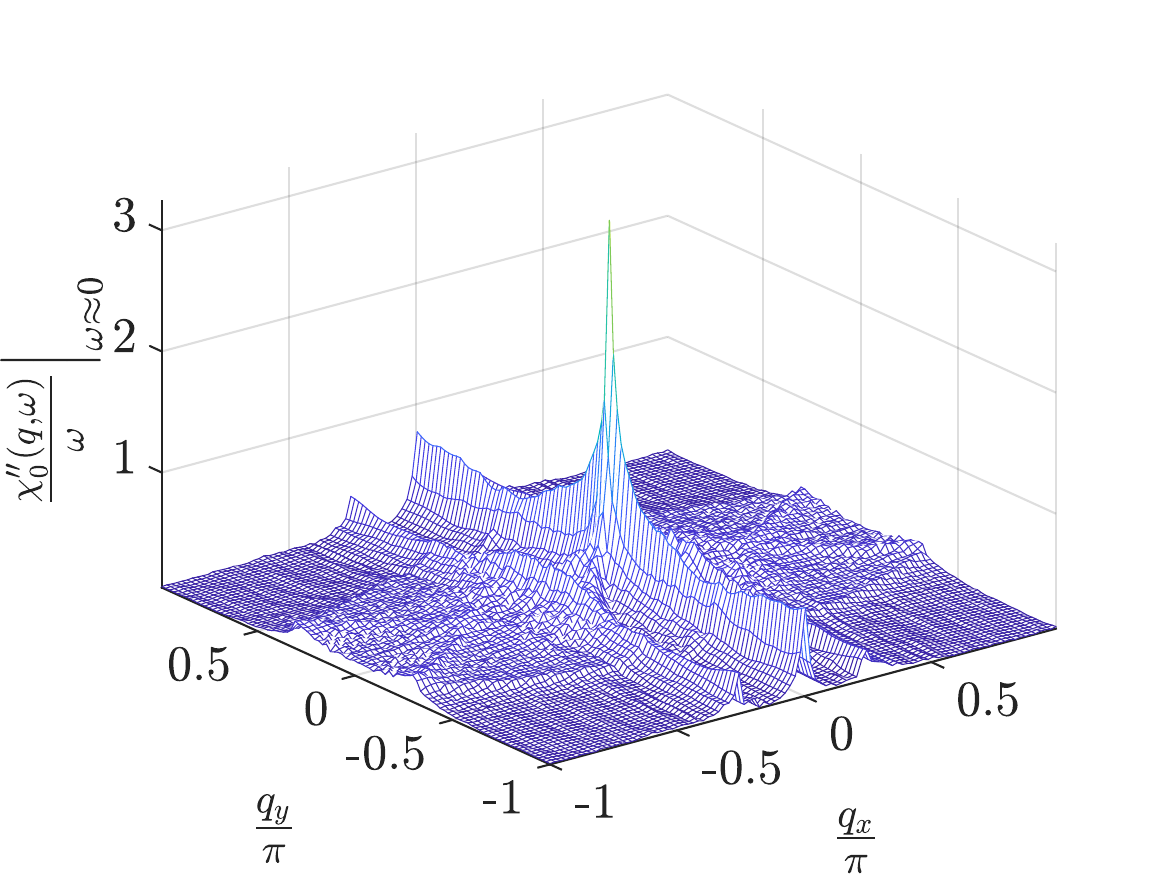}
		\put (0,70) {\bf a)}
		\end{overpic} \qquad
		\begin{overpic}[width=0.45\textwidth]{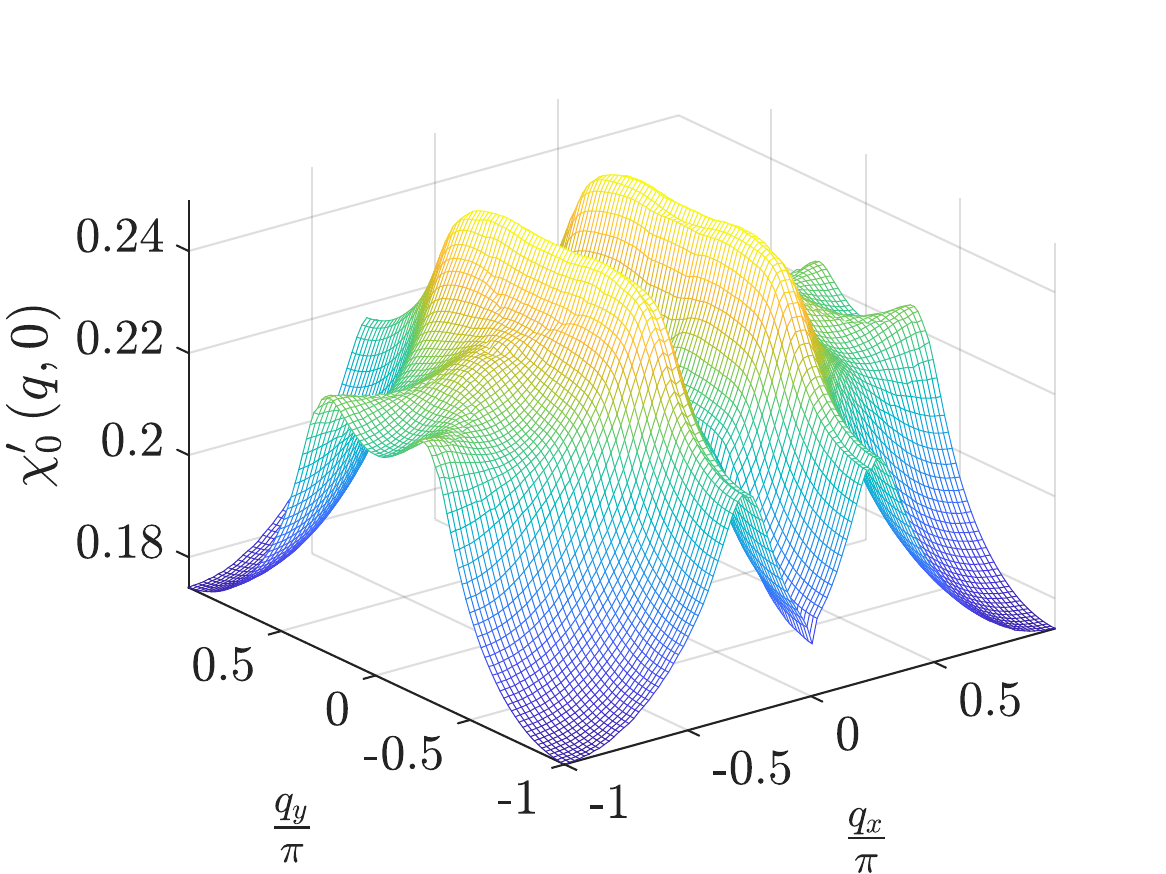}
		\put (0,70) {\bf b)}
		\end{overpic} \\
		\begin{overpic}[width=0.45\textwidth]{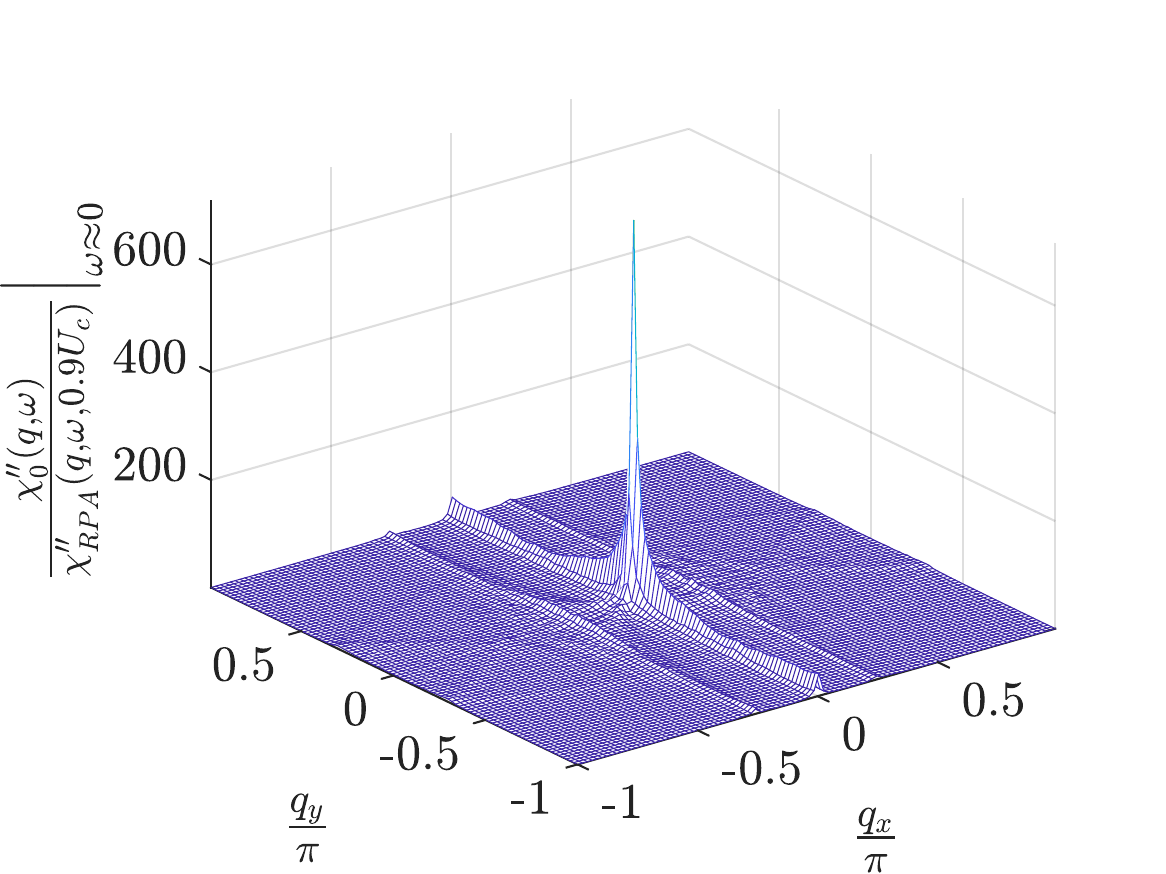}
		\put (0,70) {\bf c)}
		\end{overpic} \qquad
		\begin{overpic}[width=0.45\textwidth]{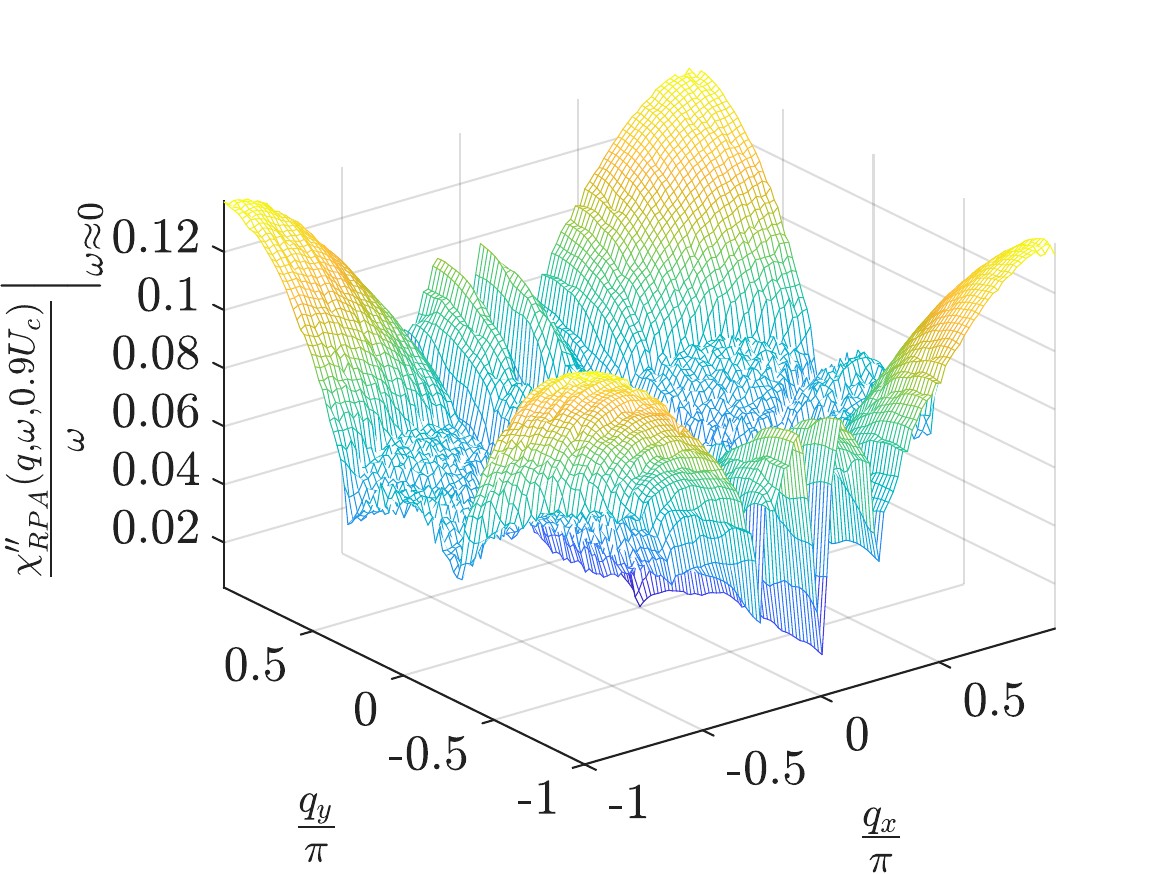}
		\put (0,70) {\bf d)}
		\end{overpic} 
	\caption{The transverse susceptibility of $\kappa$-Br with an isotropic superconducting gap, at $T=0.98T_c$, with $\omega =  10^{-3} t$. The imaginary (a) and real (b) parts of the susceptibility in the absence of spin fluctuations ($U=0$), are both enhanced around $\bm{q}=\bm{0}$. In the case of the imaginary part, the divergence near  $\bm{q}=\bm{0}$ is responsible for the Hebel-Slichter peak. The denominator of the RPA dressed susceptibility (c) is shown for an interaction strength of $0.9U_c$, for which the Hebel-Slichter peak is strongly suppressed. The denominator of the RPA shows a peak that grows noticeably as the interaction strength increases, masking the divergence of the bare imaginary part, as can be seen in (d), which shows the imaginary part of the RPA dressed susceptibility. Interestingly, though features away from $\bm{q}=\bm{0}$ are significantly enhanced, beyond the magnitude of the peak in the bare susceptibility, for both bands, these features do not contribute to the Hebel-Slichter peak, which is strongly suppressed at this interaction strength.}
	\label{Chi_sRPA}
\end{figure*}

In order to better understand the origin of this suppression of the peak we examine, in Fig. \ref{Chi_sRPA}, the properties of the transverse susceptibility for the model of $\kappa$-Br close to $T_c$. In the absence of spin fluctuations, the Hebel-Slichter peak results from a peak in the imaginary part of the susceptibility at $\bm{q}=\bm{0}$, which is present in both bands and only at higher temperatures. In the presence of antiferromagnetic fluctuations, this peak is suppressed due to both the broad maximum of the real part of the susceptibility as well as the influence of the peaked imaginary part in the denominator of the RPA-dressed susceptibility. 

As can be seen in  Fig. \ref{Chi_sRPA}d, the imaginary part of the dressed susceptibility shows no divergence, due to the cancellation of peaks in the bare imaginary susceptibility (Fig. \ref{Chi_sRPA}a) and the denominator of the dressed susceptibility (Fig. \ref{Chi_sRPA}c). While the antiferromagnetic fluctuations described by the RPA introduce some considerable structure in the susceptibility away from $\bm{q}=\bm{0}$, arising from the structure of the real part of the bare susceptibility, these features do not protect the Hebel-Slichter peak from suppression.

\begin{figure}
\centering 
	\includegraphics[width=0.45\textwidth]{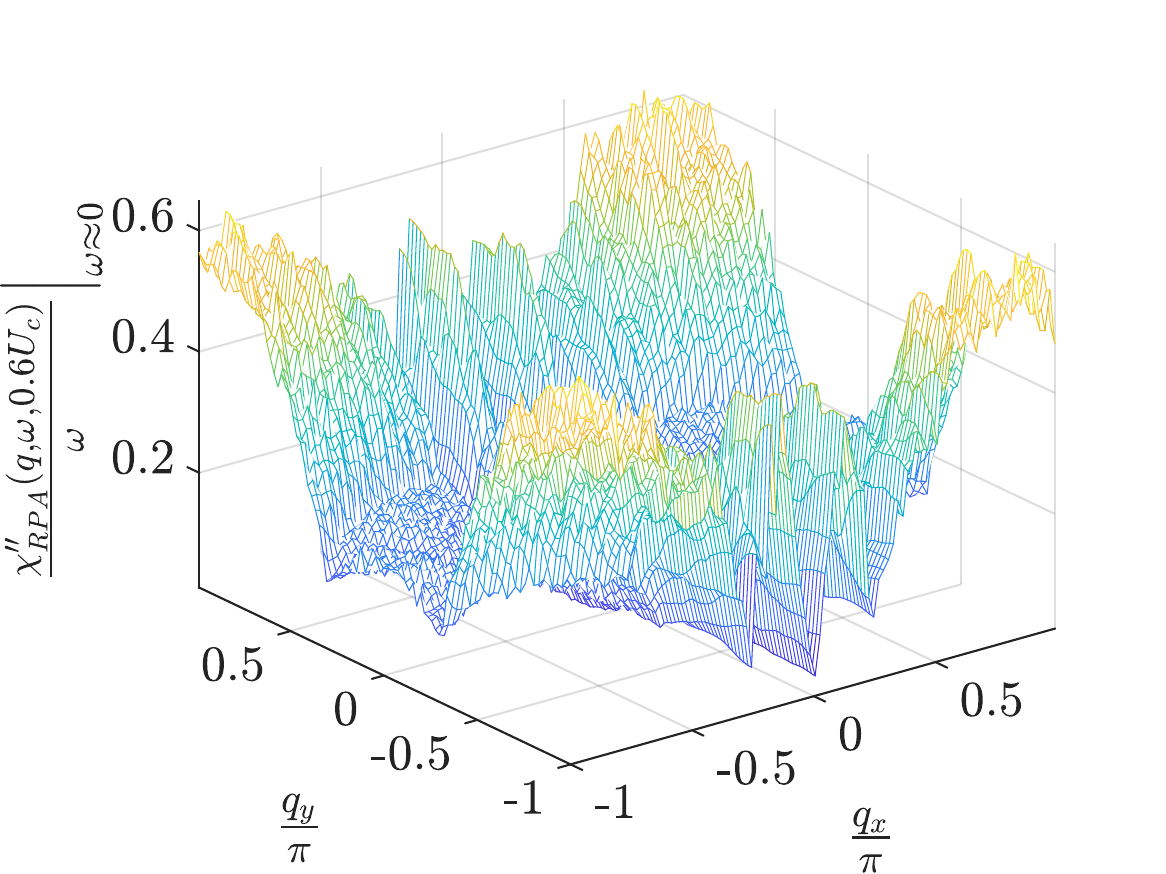}
	\caption{The RPA-dressed transverse susceptibility of $\kappa$-Br with an isotropic superconducting gap, at $T=0.98T_c$ and $\omega =  10^{-3} t$, with an intermediate interaction strength of $0.6U_c$, for which the Hebel-Slichter peak is only partially suppressed.  While the susceptibility shares many of the same features as the dressed susceptibility in Fig. \ref{Chi_sRPA}, the magnitude of the susceptibility is considerably greater, leading to the non-vanishing peak in $1/T_1T$.}
	\label{Chi_sRPA2}
\end{figure}

In Fig. \ref{Chi_sRPA2}, we examine the dressed susceptibility at an intermediate interaction strength, not sufficient to suppress the Hebel-Slichter peak entirely. In this case, there is clearly still a large enhancement of the imaginary part of the susceptibility, though the prominent peak around $\bm{q}=\bm{0}$ is no longer present. As the interaction strength increases, the overall magnitude of the susceptibility decreases further, ultimately suppressing the peak in $1/T_1T$ entirely. Additionally, the features away from $\bm{q}=\bm{0}$, while greater in magnitude, have not yet reached the definition seen in Fig. \ref{Chi_sRPA}, highlighting that both the suppression of the $\bm{q}=\bm{0}$ peak, and therefore the Hebel-Slichter peak, and the enhancement of the other features, arise due to the influence of the spin fluctuations.

\begin{figure}
\centering
		\begin{overpic}[width=0.45\textwidth]{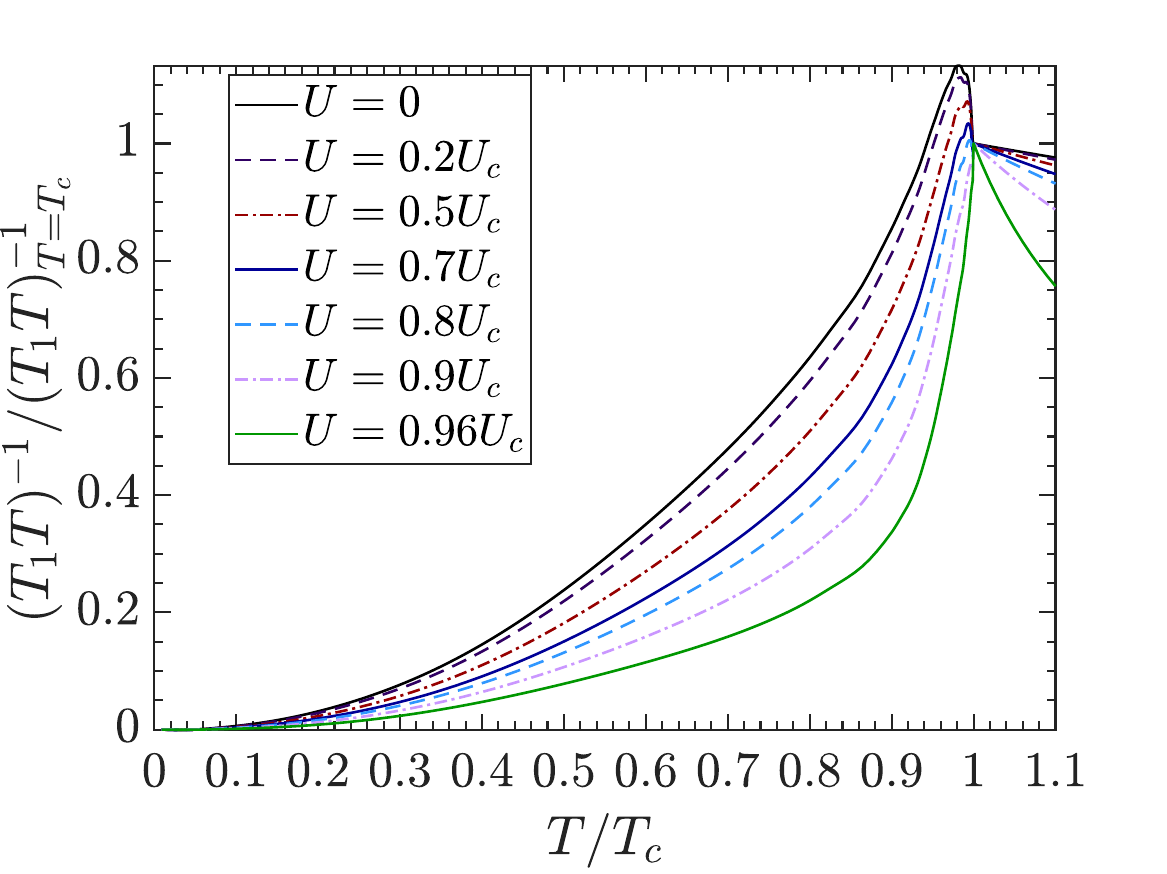}
		\put (0,70) {\bf a)}
		\end{overpic} \\
		\begin{overpic}[width=0.45\textwidth]{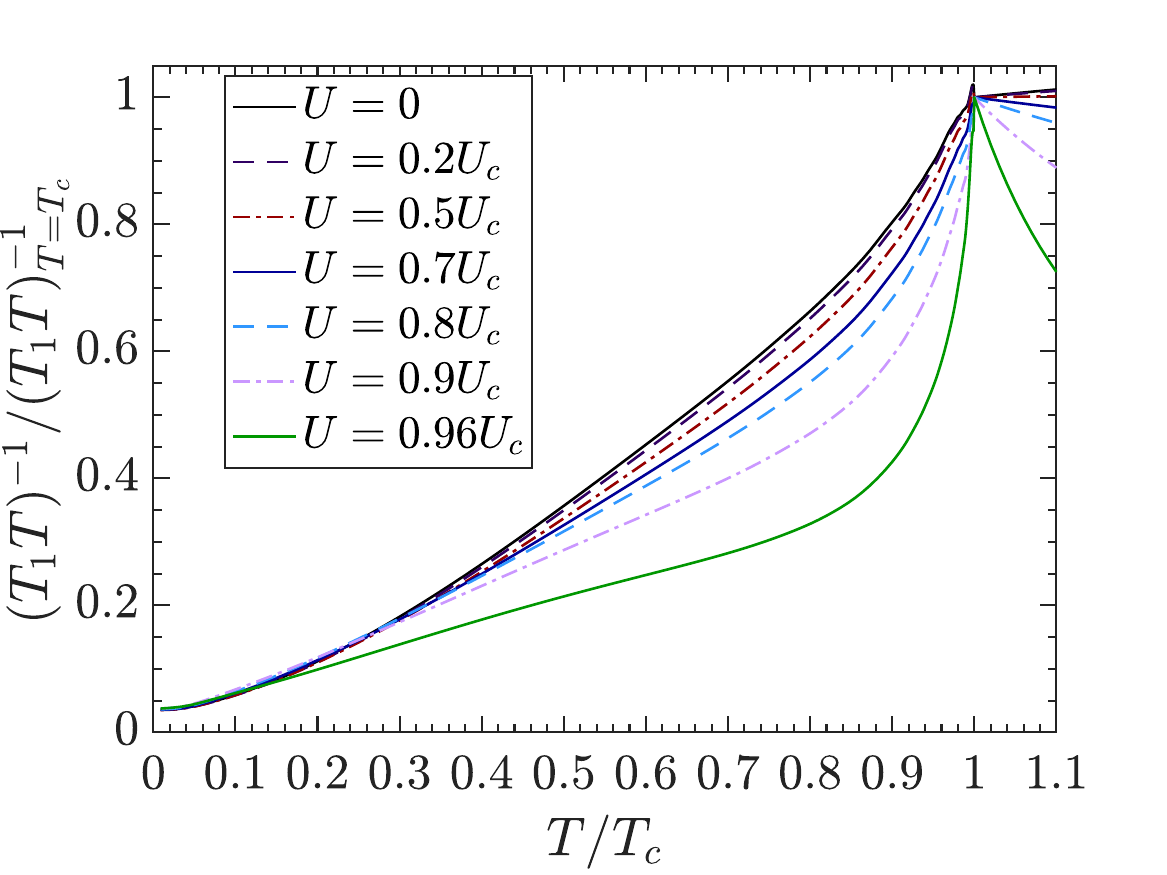}
		\put (0,70) {\bf b)}
		\end{overpic} 
	\caption{The temperature dependence of $1/T_1T$, and suppression of the Hebel-Slichter-like peak resulting from a gap with accidental nodes, with increasing interaction strength in (a) the orthorhombic model with $t_x=0.4t_y$, and (b) a model for $\kappa$-Br. The Hebel-Slichter like peak due to the accidental nodes is clearly evident for the orthorhombic model at weak interaction strengths, and vanishes in the presence of strong antiferromagnetic fluctuations. While the peak only disappears for $U\geq 0.8U_c$, the narrowing and suppression of the peak at lower interaction strengths may be sufficient to disguise the Hebel-Slichter peak in experiments. For the model of $\kappa$-Br, the peak is considerably smaller and narrower, and therefore less likely to be clearly resolved even in the absence of strong antiferromagnetic fluctuations. However, the influence of the accidental nodes is still clear at weak interaction strengths, where the relaxation rate decreases considerably less rapidly as the temperature is reduced below $T_c$. This effect is also suppressed as the interaction strength increases, but may in principle be examined experimentally by the application of pressure, which reduces the effective interaction strength in these materials.  At low temperatures, the temperature dependence of $1/T_1T$ for both models is again not qualitatively altered by the introduction of spin fluctuations via the RPA. In these calculations, $\Delta_0/2=2.5k_BT_c$, $\eta = 5\times 10^{-3} t$, and $\omega =  10^{-3} t$,  while $U_c\sim 11.5 t$ for the orthorhombic model and $U_c\sim 9.6 t$ for the model of $\kappa$-Br. }
	\label{U_aRPA}
\end{figure}

We wish also to understand how this suppression influences the Hebel-Slichter like peak expected in superconductors with accidental nodes \cite{Cavanagh2018}. Fig. \ref{U_aRPA} displays the suppression of the Hebel-Slichter-like peak for the orthorhombic and $\kappa$-Br  models with d$_{x^2-y^2}$-wave and d$_{xy}$-wave superconducting gaps, respectively, each with accidental nodes. The peak is suppressed in the same manner as in the previous case, though much more rapidly with increasing interaction strength.

The low temperature behavior for the gaps with accidental nodes in Fig. \ref{U_aRPA} does not show the exponential suppression of quasiparticle states seen for the isotropic gap, but is again qualitatively unchanged by the increasing interaction strength. Interestingly, the relaxation rate changes much more dramatically as $U\rightarrow U_c$ for the more realistic $\kappa$-Br model, most likely due to a singularity in the density of states, which is much closer to the Fermi energy than for the orthorhombic model. It may be necessary, in general, to examine the low temperature behavior of $1/T_1T$, and not just the presence or absence of a peak near $T_c$ to infer the superconducting gap symmetry.

\begin{figure}
\centering
	\begin{overpic}[width=0.45\textwidth]{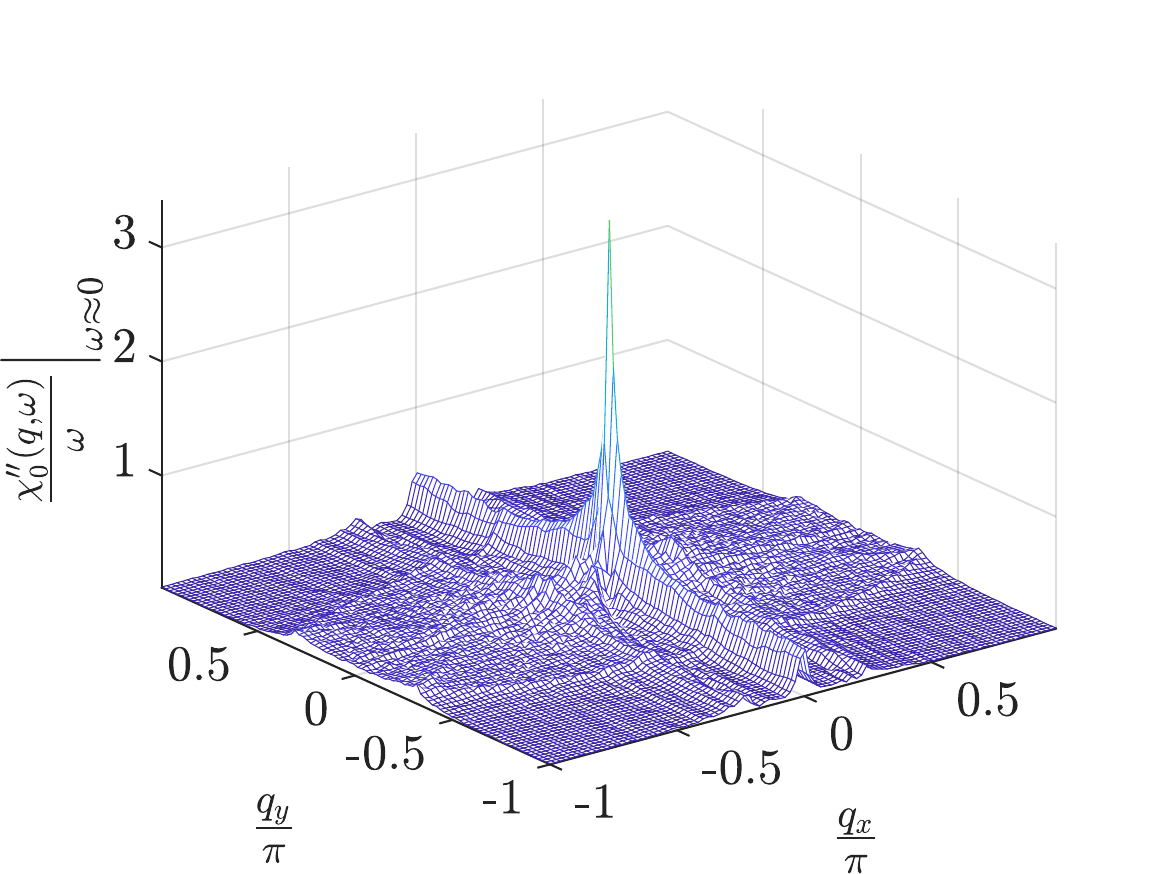}
		\put (0,70) {\bf a)}
		\end{overpic} \\ 
	\begin{overpic}[width=0.45\textwidth]{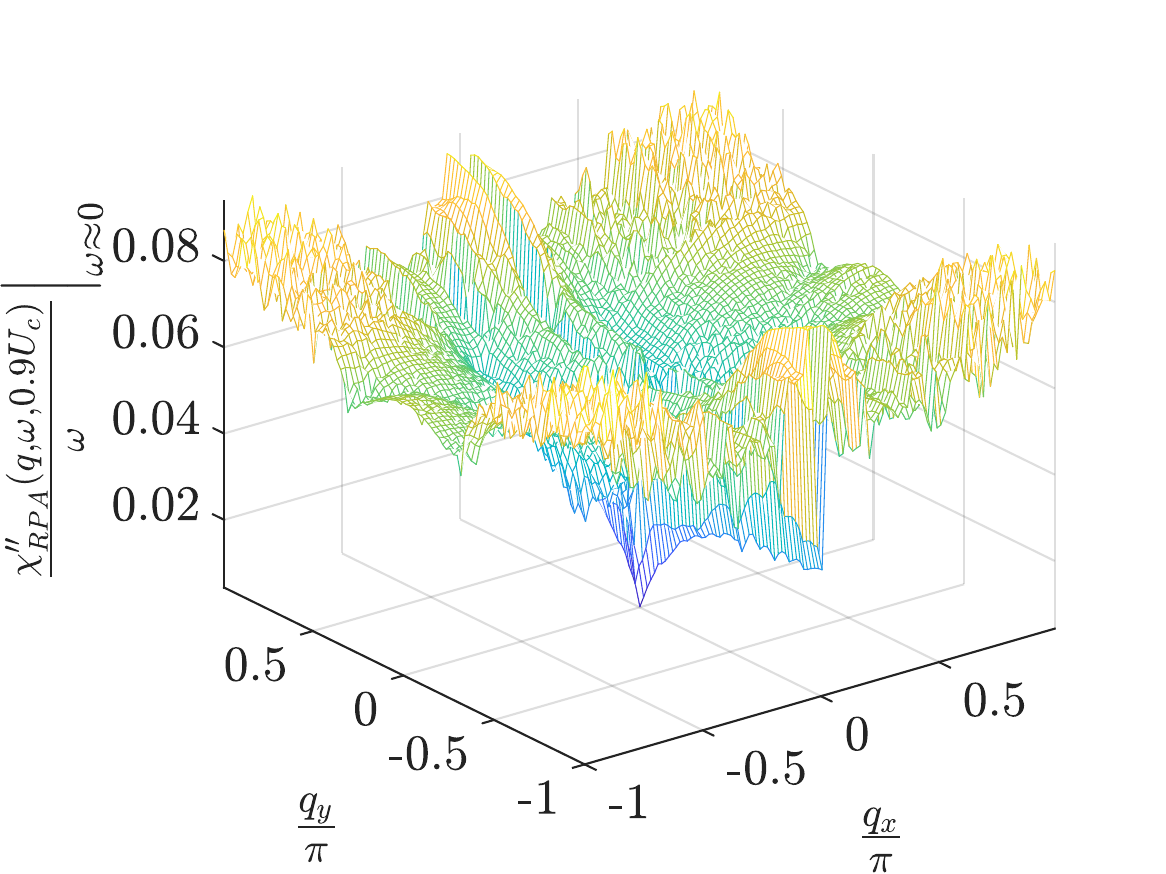}
		\put (0,70) {\bf b)}
		\end{overpic} 
	\caption{The bare (a) and RPA-dressed (b) transverse susceptibility of $\kappa$-Br with a superconducting gap with accidental nodes, at $T=0.98T_c$ and$\omega =  10^{-3} t$, are qualitatively the same as those for an isotropic gap. The imaginary part of the susceptibilities in the absence of antiferromagnetic fluctuations again diverges near $\bm{q}=\bm{0}$, and the reduced magnitude of the peak in $1/T_1T$ relative to that for an isotropic gap is directly related to the reduced width of the peak in $\chi_0''$. The RPA dressed susceptibility shows no such divergence, consistent with the absence of a peak in $1/T_1T$. Additionally, the dressed susceptibility varies considerably less across the Brillouin zone than in the presence of an isotropic gap, which allows the suppression of the susceptibility at $\bm{q}=\bm{0}$ to be seen far more clearly.}
	\label{Chi_aRPA}
\end{figure}

In Fig. \ref{Chi_aRPA}, we again examine the origin of the peak suppression for the gap with accidental nodes in $\kappa$-Br, finding a situation at high temperatures ($T=0.98T_c$) that is qualitatively the same as the nodeless gap. The peak in $1/T_1T$ is caused by a peak in $\chi''\left(\bm{q},\omega\right)/\omega$ near $\bm{q}=\bm{0}$, which is suppressed by the RPA as the interaction increases. Additionally, the RPA-dressed susceptibilities in both cases are qualitatively similar, differing only in the magnitude of variation in the susceptibility across the Brillouin zone, despite the significant reduction in the magnitude of the peak in $1/T_1T$. Notably, the suppression of the susceptibility near $\bm{q}=\bm{0}$ by antiferromagnetic fluctuations is more clearly apparent due to the smaller magnitude and reduced variation of the susceptibility. This further solidifies the similarities between the two gap functions, despite the presence of line nodes in the second case, and the corresponding alteration of the density of states.

\begin{figure}
\centering 
	\includegraphics[width=0.45\textwidth]{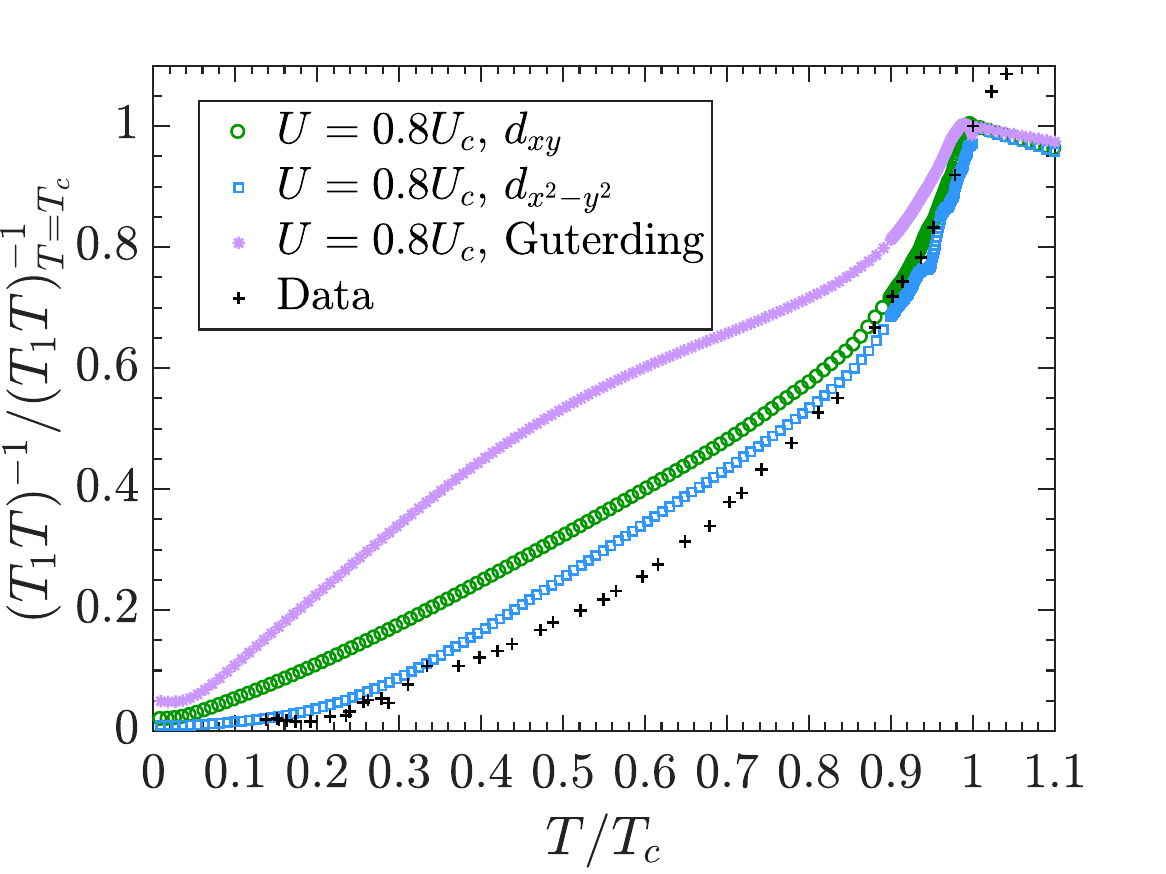}
	\caption{Comparison between results for the dimer model of $\kappa$-Br for gaps with accidental ($d_{xy}$) and with symmetry required ($d_{x^2-y^2}$) nodes, as well as a nodal $s+d_{xy}$-wave gap proposed by Guterding \emph{et al}. \cite{Guterding2016} (`Guterding'), with experimental data from Ref. \cite{Kanoda1996}. The spin-fluctuation strength here is sufficient to suppress the Hebel-Slichter like peak for the $d_{xy}$-wave gap, and while both $d_{xy}$ and $d_{x^2-y^2}$ gaps agree well with the data immediately below $T_c$ the $d_{x^2-y^2}$-wave state is in noticeably better agreement at low temperatures. The $s+d_{xy}$-wave gap proposed by Guterding \emph{et al}. based on scanning tunneling spectroscopy experiments does not fit the NMR data well. In these calculations, $\eta = 5\times 10^{-3} t$, $\omega =  10^{-3} t$, $\Delta_0/2=2.5k_BT_c$, and $U_c\sim 9.6t$ for $\kappa$-Br regardless of superconducting gap.}
	\label{k_data}
\end{figure}

Finally, we make comparison between our results for the model of $\kappa$-Br and experimental data \cite{Kanoda1996}, in Fig. \ref{k_data}. We find that sufficiently strong spin-fluctuations suppress the Hebel-Slichter like peak for a $d_{xy}$-wave gap with accidental nodes, in good agreement with experiment, though the low temperature behavior is less consistent. At low temperatures, our results predict a relaxation rate that increases far more rapidly with temperature than is observed experimentally. For comparison, we also consider a `$d_{x^2-y^2}$-wave' gap with symmetry required nodes, and find much closer agreement with the experimental data at low temperatures. 
For completeness, we also consider a recently proposed $s+d_{xy}$-wave gap, found to be in agreement with some scanning tunnelling spectroscopy measurements \cite{Guterding2016a, Guterding2016}. Again, in this case the Hebel-Slichter like peak is suppressed by the spin-fluctuations, but we find the calculated relaxation rate to be in considerably worse agreement with the NMR data than the simple $d_{xy}$-wave gap. 
A more definitive test of the gap structure may be performed by the application of pressure, which for the BEDT-TTF-based superconductors reduces the effective interaction strength $U$. Applied pressure would then restore the Hebel-Slichter like peak for the $d_{xy}$-wave gap but affect the relaxation rate for the $d_{x^2-y^2}$-wave gap less significantly.

\begin{figure}
\centering 
	\includegraphics[width=0.45\textwidth]{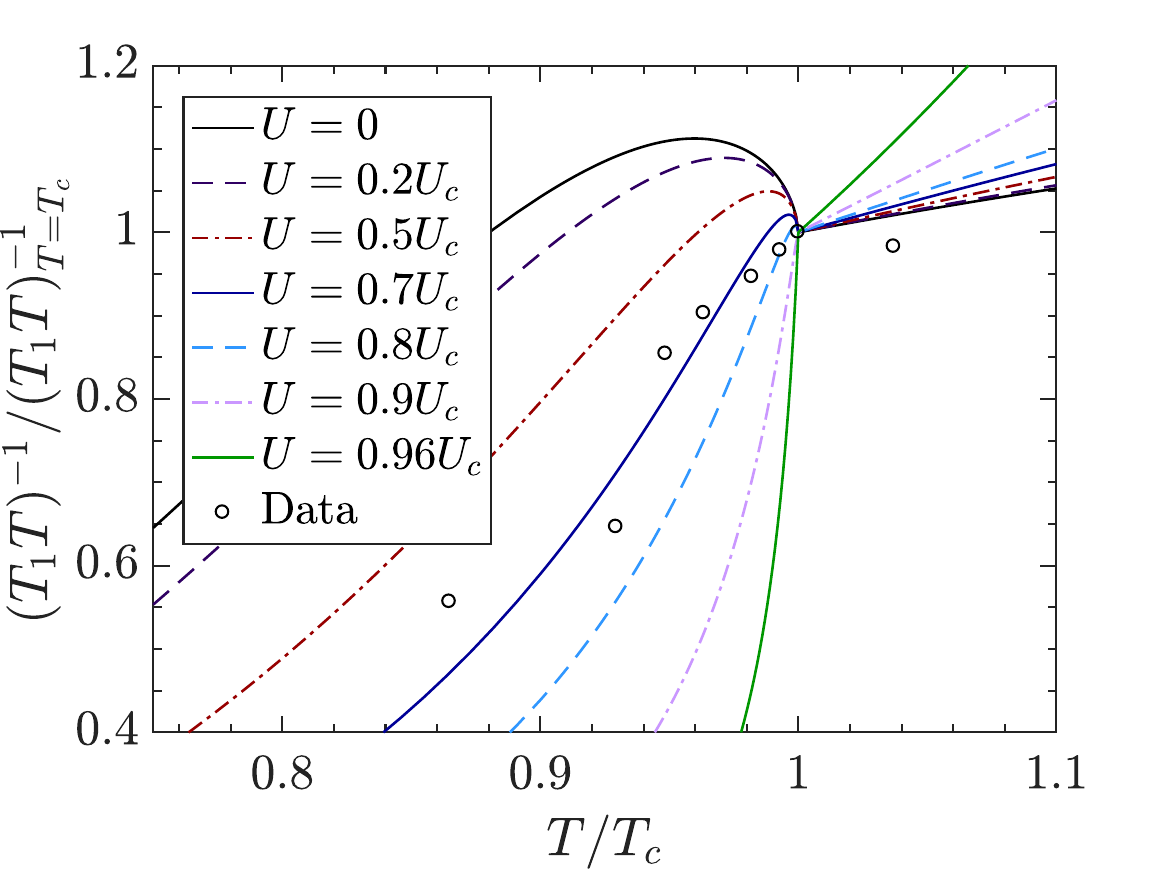}
	\caption{Comparison between numerical results for the two-orbital iron pnictide model with $s_{\pm}$-wave gap and experimental data for LaFeAsO$_{1-x}$F$_x$, at $x=0.06$, reported in Ref. \cite{Oka2012}. Increasing the strength of the spin-fluctuation suppresses the Hebel-Slichter peak in line with the experimental data for $U\sim 0.7U_c$. In these calculations, $\eta = 5\times 10^{-3} t$, $\omega =  10^{-3} t$, and $\Delta_0/2=2.5k_BT_c$, in agreement with Ref. \cite{Oka2012}, and $U_c\sim 9.3|t_1|$.}
	\label{FeAs_data}
\end{figure}

We make comparison between the relaxation rate calculated for the two-orbital iron pnictide model, with $s_{\pm}$-wave superconductivity, and experimental data for LaFeAsO$_{1-x}$F$_x$ in Fig. \ref{FeAs_data}. At low temperatures, complicated variations in the gap magnitude on the multiple bands lead to a rich temperature dependence of $1/T_1T$ in LaFeAsO$_{1-x}$F$_x$. As we are primarily concerned with the behavior of the Hebel-Slichter peak, we restrict our analysis to the region immediately below the critical temperature, $T\geq 0.8T_c$. We again find that strong spin-fluctuations remove the Hebel-Slichter peak for LaFeAsO$_{1-x}$F$_x$, in agreement with experiment.

\section{Conclusions}

We have found that, for all model bandstructures we consider, sufficiently strong antiferromagnetic spin fluctuations suppress the Hebel-Slichter peak in a fully gapped superconductor, and the similar peak found for gaps with accidental nodes. Even when the peak is suppressed by the spin fluctuations, near $U/U_c \approx 1$, the  low temperature behavior of the nuclear magnetic relaxation rate remains qualitatively unchanged by the interactions. This is because the magnitude of both the real and imaginary parts of the susceptibility near $\bm{q}=0$ decreases as the temperature is lowered. And so, just as the Hebel-Slichter peak is only evident near $T=T_c$, the influence of the spin fluctuations is less significant at low temperatures. 


In the organic superconductors, the application of pressure can be used to decrease the effective interaction strength, which will increase the magnitude of any peak in $1/T_1T$. Therefore, we propose an additional experimental probe of the superconducting gap in these materials, by measuring the temperature and pressure (and therefore $U/U_c$) dependence of $1/T_1T$ to give further insight into the gap symmetry. In particular, for a nodeless gap, or one with accidental nodes, a Hebel-Slichter peak should appear as pressure is increased. For a gap with symmetry required nodes, no such peak will emerge under pressure. 

\begin{acknowledgments}We thank  Qiang-Hua Wang and J. Wosnitza for helpful conversations, as well as Kazushi Kanoda and Guo-Qing Zheng for access to experimental data.
 This work was supported by the Australian Research Council (Grant No.
 DP180101483) and by an Australian Government Research
 Training Program Scholarship. 
\end{acknowledgments}

\maketitle
\bibliographystyle{unsrt}
\bibliography{Ongoing_Bib}
\end{document}